# Low-temperature electron-phonon interaction of quantum emitters in hexagonal Boron Nitride

Gabriele Grosso[1,2,*,‡], Hyowon Moon[3,‡], Christopher J. Ciccarino[4], Johannes Flick[4,†], Noah Mendelson[5], Milos Toth[5], Igor Aharonovich[5], Prineha Narang[4] and Dirk R. Englund[3]

1. Photonics Initiative, Advanced Science Research Center, City University of New York, New York, NY, USA. 2. Physics Program, Graduate Center, City University of New York, New York, NY, USA. 3. Department of Electrical Engineering and Computer Science, Massachusetts Institute of Technology, Cambridge, MA, USA. 4. John A. Paulson School of Engineering and Applied Sciences, Harvard University, Cambridge, MA, USA. 5. School of Mathematical and Physical Sciences, University of Technology Sydney, Ultimo, New South Wales 2007, Australia.
* Corresponding author: ggrosso@gc.cuny.edu
‡ Equal contribution
† Present address: Center for Computational Quantum Physics, Flatiron Institute, New York, NY, USA

Quantum emitters based on atomic defects in layered hexagonal Boron Nitride (hBN) have emerged as promising solid state "artificial atoms" with atom-like photophysical and quantum optoelectronic properties. Similar to other atom-like emitters, defect-phonon coupling in hBN governs the characteristic single-photon emission and provides an opportunity to investigate the atomic and electronic structure of emitters as well as the coupling of their spin- and charge-dependent electronic states to phonons. Here, we investigate these questions using photoluminescence excitation (PLE) experiments at $T=4\ K$ on single photon emitters in multilayer hBN grown by chemical vapor deposition. By scanning up to 250 meV from the zero phonon line (ZPL), we can precisely measure the emitter's coupling efficiency to different phonon modes. Our results show that excitation mediated by the absorption of one in-plane optical phonon increases the emitter absorption probability ten-fold compared to that mediated by acoustic or out-of-plane optical phonons. We compare these measurements against theoretical predictions by first-principles density-functional theory of four defect candidates, for which we calculate prevalent charge states and their spin-dependent coupling to bulk and local phonon modes. Our work illuminates the phonon-coupled dynamics in hBN quantum emitters at cryogenic temperature, with implications more generally for mesoscopic quantum emitter systems in 2D materials and represents possible applications in solid-state quantum technologies.

"Artificial atoms" based on point-like imperfections in solids are emerging as excellent systems for quantum applications ranging from sensing and networks to information processing.[1–3] Such emitters were recently observed in two-dimensional (2D) crystals of hexagonal Boron Nitride (hBN), a layered wide-gap insulator in which structural defects can create energy levels confined within the band gap[4,5]. Many key functionalities have been reported including room temperature operation,[4] high brightness and quantum efficiency,[6,7] spin structure,[8] narrow linewidth,[9] spectral tuning via applied strain[7] and the Stark effect.[10–12] However, despite electron microscopy-based imaging[13] and theoretical electronic structure studies[5,14], the atomic structure of observed hBN defects is still under debate. Moreover, the problems of large spectral distribution[15] and severe spectral wandering[16] limit applications.

The emission spectrum of a point defect emitter strongly depends on its interaction with phonon modes. Combined with electronic structure models, emission spectroscopy therefore gives important insight into the structural properties of the emitter. Structural defects in solids can be considered as molecules with vibrational states coupled to electronic states, as captured in the Huang-Rhys model[17] illustrated in Figure 1a. Therefore, for each electronic configuration (i.e. ground or excited states in Figure 1a), emission and absorption can be mediated by phonons. A typical photoluminescence spectrum from hBN emitters consists of a zero-phonon line as illustrated by the $n=n^*$ transitions in Figure 1a, in addition to lower-energy phonon-assisted sidebands for which $n>n^*$. The absorption efficiency shows a blue-detuned phonon sideband when the excitation source matches transitions with $n<n^*$.

The coupling of hBN electronic states with optical phonons creates distinctive phonon sidebands that are visible in the emission spectrum at room and low temperature.[15] Recent studies have indeed shown a dependence of the external quantum efficiency of hBN quantum emitters on coupling to hBN optical phonons[18–20]. However, because these measurements were performed at room temperature and with limited excitation wavelengths, they cannot resolve the coupling to different optical phonon modes (TO and LO) and resonances, or coupling to relevant acoustic phonon modes. Moreover, previous experimental works did not connect spectroscopy to individual localized phonon resonances of candidate hBN atomic defects.

In this *Letter*, we study the absorption spectrum of quantum emitters in hBN at cryogenic temperature by photoluminescence excitation experiments with meV-scale resolution. These cryogenic, high-resolution measurements enable quantitative comparisons to theoretical predictions of defect-phonon coupling, including relevant lattice phonons (acoustic and optical bands) and localized phonons specific to the most likely atomic defect structures.[5,21] Our theoretical calculations are based on density functional theory for different charge states of atomic defect structures and include, for the first time for hBN quantum emitters, the effects of spin polarization on phonon sideband coupling (see Methods). The experiments are performed on emitters in few-layer hBN films produced via chemical vapor deposition (CVD) on copper, which are then transferred onto $Si/SiO_2$ substrate for further analysis[10] (see Methods). Thin-layer hBN films are desirable as the emission properties can be tailored during growth,[22–24] and present a platform amenable to integration with van-der waals heterostructures.[25] This CVD grown sample has a mean density of quantum emitters of ~2 /$\mu m^2$ with >85% of ZPLs centered around 2.14 eV, suggesting that one type of structural defect is predominant in this sample.[10]

We performed optical experiments in a closed-cycle cryostat with top-mounted confocal microscope, using narrow-band (< 1 MHz) tunable laser excitation with a diffraction-limited spot size of 300 nm for 532 nm excitation, and free-space coupling of emission to a spectrometer or single photon counting modules (see Methods). The photoluminescence maps in Figure 1c and Figure 1d compare the hBN emission in the same characteristic 10 μm x 10 μm sample region for excitation energy of 2.13 eV and 2.33 eV. The first excitation frequency is chosen to be near the center of the characteristic ZPL energy distribution of the sample, while the second frequency is chosen to be ~ 0.19 meV above, as illustrated in Figure 1b. The spatial and intensity distribution of the emitteres changes dramatically due to frequency dependent absorption of the emitters.[18] Specifically, we observe that in Figure 1c, the ZPL energy of the very few visible emitters mostly falls in the range 1.89 - 1.97 eV, with a mean laser detuning around 160 meV. In contrast, many emitters appear with energy up to 2.17 eV in Figure 1d, in agreement with previous characterization of this sample.[10] This change of photoluminescence as a function of the excitation energy is explained by the relative coupling strength with different phonon modes that, as discussed below, is more efficient for detuning in the range of 150-200 meV.

We consider now spectroscopy on a single exemplary emitter "*E2*" indicated in Figure 1b. Figure 2a plots the emission spectra (in logarithmic color scale) as the excitation energy is tuned from $\omega_{ZPL}$ + 60 meV to $\omega_{ZPL}$ + 252 meV, where $\omega_{ZPL}$ is the zero phonon line energy of emitter *E2*. For each excitation energy $\omega_i$, spectra are normalized by the laser excitation power $P_i$, which fluctuated by ~ 80 µW around a mean of 120 µW (measured before the objective lens). The typical saturation power for emitters in this sample is $P_{sat}$ ~ 60 µW near the absorption resonance (Supplementary Figure 1S). To include this effect, each spectrum is normalized by a factor $(P_i + P_{sat})/P_i$. To mitigate spectral wondering (Supplementary Figure 2S), we integrate spectra over one minute and obtain the laser detuning ($\Delta E_{PLE} = \omega_i - \omega_{ZPL}$) from the mean of the ZPL peak for each spectrum. The emitter fluorescence intensity rises by more than an order of magnitude when the laser energy is tuned around $\Delta E_{PLE}$ = 165 meV compared to detuning lower than 150 meV (Figure 2a), in agreement with previous studies.[18,26] Figure 2b compares the emission spectra at two different detunings, $\Delta E_1$ = 165 meV and $\Delta E_2$ = 120 meV (white dashed lines in Figure 2a). Photon antibunching in the correlation histogram in Figure 2c, well below 0.5 (fitted by $g^{(2)}(0)$ = 0.11 ) assures that this is a single-photon emitter.

Figure 3 plots the photoluminescence (PL) intensity as a function of $\Delta E_{PL} = \omega_{ZPL} - \omega_{PL}$ (for excitation energy 2.33 eV) and normalized PLE intensity (integrating over the ZPL) for emitter *E1*, *E2*, and *E3* (Figure 1b) with similar ZPL energies between 2.10 and 2.16 eV.[15] In the detuning range $\Delta E_{PLE}$ =50 - 150 meV, PLE experiment shows low absorption with weak peaks: a peak at 85 meV matching the energy of *two* out-of-plane acoustic phonons at the M point (*ZA(M)*) that each have ~ 42 meV as measured by second-order Raman spectroscopy;[27,28] and a peak at 120 meV that we tentatively ascribe to *ZA(K) (40 meV)* and *ZO(K) (80 meV)* phonons at the K point.[27,28] The resonance around 120 meV is not observed in the emission spectra.

The PLE spectra for all emitters show three pronounced peaks near $\Delta E_{PLE}$ = 157 meV, 168 meV, and 183 meV, which are matched by PL peaks at $\Delta E_{PL}$=159 meV and 167 meV (our PL spectra were cut off at 182 meV). These experimental data are in good agreement with the results of Raman spectroscopy and ab initio calculations of three phonon lines (*LO $K_{1,2}$, LO $\Gamma_{5+}$*, and *LO overbending)* at 158 meV, 169 meV, and 185 meV.[28,29] The peak intensities suggest varying coupling strengths between these hBN phonons and the defect: the 158 meV peak is similar in intensity to the others even though the *LO $K_{1,2}$* has almost twice the density of states.[30] Given the energy match between our observation and theoretical predictions on phonon modes in hBN, we tentatively attribute the enhanced absorption observed in the range $\Delta E_{PLE}$= 150 - 200 meV to the mediation of one in-plane longitudinal optical phonon. This energy range is in agreement with a recent study on the absorption and emission dipole orientation of hBN quantum emitters.[31]

The absorption spectrum of emitters with ZPL energy below 2.10 eV, thus away from the typical mean value of the sample, show significant differences. In these cases, the different defect-phonon coupling suggests that a different atomic structures could be responsible for the quantum emission. Examples of PLE and PL spectra for emitters with ZPL energy at 2.03 eV and 2.08 eV are shown in Supplementary Figure 3S and Figure 4S, respectively.

While much theoretical work has focused on correlating ZPL energies with experimental findings[12,21,32], geometry shifts and therefore defect-phonon coupling provide an additional fingerprint useful to identifying defect candidates that could correspond to the emitters seen in experiments. We now attempt to associate the typical emitter of our sample with a specific defect structure by comparing our meV-scale resolution absorption measurements to theoretical predictions of defect-phonon coupling. We calculate the ground and excited state defect geometries and phonon coupling via first-principles density-functional theory[32,33] with particular spin-resolved transitions in four potential defect candidates: $C_BV_N$, with transitions involving singlet and triplet ground states, and the two different spin-channels (spin-majority/spin-minority) in $N_BV_N$. All defect transitions occur between localized orbitals within the band gap of hBN.[21] Nonzero dipole overlap for each of the transitions studied confirms these transitions are symmetry-allowed.

Our ΔSCF (see Methods/Supplement for theoretical and computational details) calculations find that upon excitation, significant geometry shifts occur for all the defects present. These geometry shifts correspond to weaker ZPL emission with significant emission into the phonon sidebands. In particular, the singlet $C_BV_N$ and both $N_BV_N$ transitions show this trend. Figure 4 presents the predicted partial Huang-Rhys factors associated with each of the phonon modes of the multilayered hBN systems studied. We find that much of the coupling is via low-energy, acoustic phonon modes for each of the transitions studied. Additionally, the coupling to high-energy LO phonons appears to be small for the singlet $C_BV_N$ transition, as well as the spin minority of $N_BV_N$. In contrast, the triplet transition of $C_BV_N$ and the spin-majority transition of $N_BV_N$ also show non-negligible coupling to the higher-energy phonons between 100 meV and 200 meV.

We now discuss possible hypotheses for the disparity between theory and experiment observed, both in this work and in other theoretical works; we emphasize the need for further investigation and development of new methods to capture the physics of emitters in hBN. We highlight that the defect configurations studied were constrained to sit within the plane of the hBN layer such that there was no relaxation in the z direction. Recent experimental reports[11,12] have shown defects in hBN to exhibit a linear dependence with electric field (i.e., a linear Stark effect), which suggests a permanent electric dipole moment in the *z* direction. However, out-of-plane distortions would mean that the $\sigma$ and $\pi$ bonds associated with the defect and the surrounding hBN lattice would mix, and thus the states would be of a multi-reference nature.[34] Methods such as density functional theory are based on a single Slater determinant and therefore these states would not be accurately captured; *GW* and *GW*-BSE techniques that start with DFT states would also face the same limitations. Importantly, the excited states may be of multireference nature even without an out-of-plane distortion, which would not be found using single-particle methods. Multireference schemes based on quantum chemistry (configuration interaction, coupled cluster) implementations may be useful in describing the nature of these states, however many of these methods are not currently scalable to large system sizes required to describe a defect within a solid, and in particular the associated vibrational properties.

Additionally, for the transitions studied, there was a large associated relaxation energy upon promotion of an electron to the excited orbital. This may be an artifact of the constrained 2D nature of the defect sheet. However, if this large geometry relaxation is indeed physical, the associated defect-phonon coupling method may be inadequate. The technique used across the solid-state defect community[35] relies on the assumption that the phonons of the ground and excited state are identical, which is likely not true for an excited state which takes on a significantly different geometry. Methods to address this would include a Duchinsky procedure [36,37] to relate the force constant matrix in the ground and excited state such that the correct vibrational overlaps are considered when determining the spectra. All of these options must be explored when thinking about defect transitions in hBN and will likely aid in identifying more accurately the possible defect candidates associated with experiment. We note that recent work[32] on the description of these emission sources has proposed dangling Boron bonds in the lattice, which had previously been unstudied. This and potentially other yet-to-be studied defects may also help to explain the discrepancy between theory and experiment. We hope that this candid presentation of theoretical considerations in hBN emitters will benefit the community.

In summary, we report photoluminescence excitation experiments on quantum emitters in hBN at cryogenic temperature that allows us to resolve individual absorption resonances. Absorption efficiency increases by almost an order of magnitude when the system is excited with detuning between 150 meV and 200 meV. A comparison with Raman spectroscopy and *ab initio* calculations of phonon modes in hBN suggests that this enhancement is due to absorption mediated by one in-plane longitudinal optical phonon (LO). The relative coupling efficiency of different spin transitions to phonon modes has been calculated for the first time and provides a new avenue to characterize experimental observation of defect-based quantum emitters. Our results illuminate the defect-phonon and the vibrational states of hBN emitters, indicating a new avenue to restrict the possible atomic structures associated with hBN emitters.

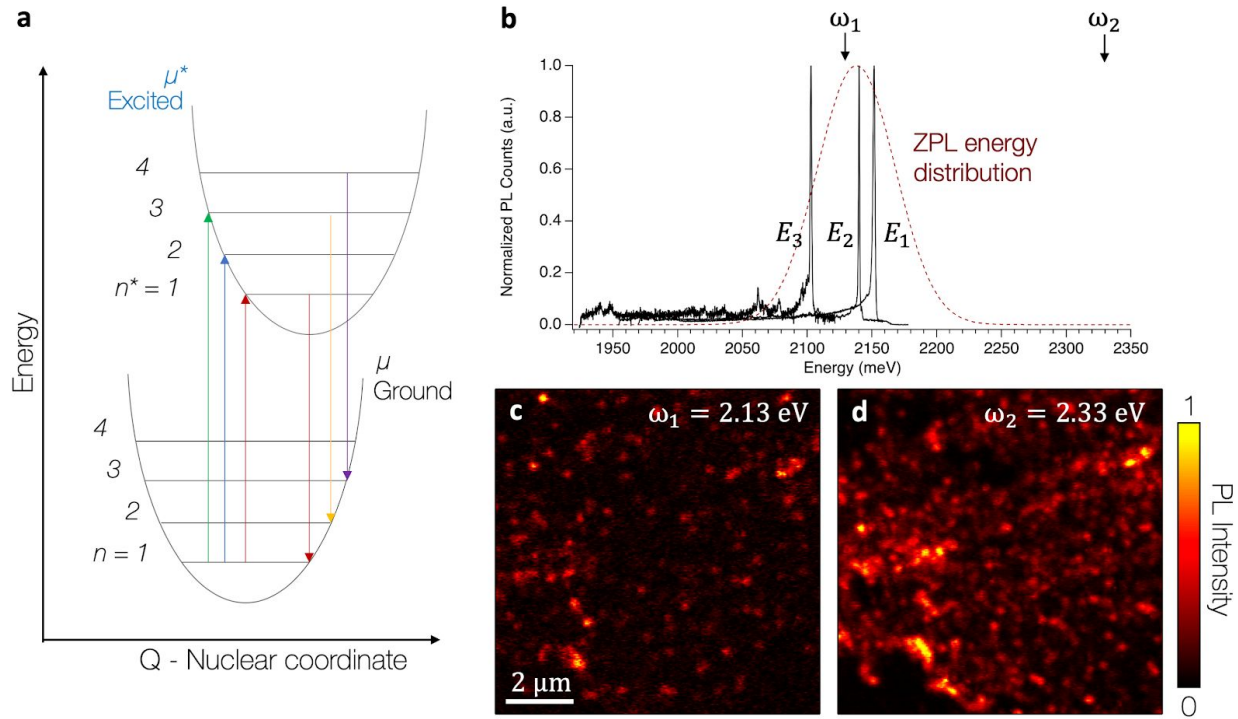

**Figure 1: Photoluminescence excitation experiment: a -** Diagram showing the Huang-Rhys Model for light absorption and emission from the vibronic states of hBN defects. Each electronic state can interact with the phonon bath that modulates the emission energy. Transition with $n = n^*$ correspond to the ZPL, while transitions with $n \neq n^*$ form the phonon sidebands. **b -** Emission spectra of three exemplary emitters (E1 - E3). The red dashed line indicates the typical energy distribution of the ZPL energy that, for this sample, is centered around 2.14 eV with a 30 meV width. **c, d -** Photoluminescence maps of the same hBN sample region excited with laser at 2.13 eV (c) and 2.33 eV (d). The two maps clearly show that the absorption of quantum emitters in hBN depends on the excitation energy. Only a few emitters are visible in (c) with energy of the ZPL in the range 1.89 - 1.97 eV. Many emitters are visible in (d) with energy up to 2.17 eV.

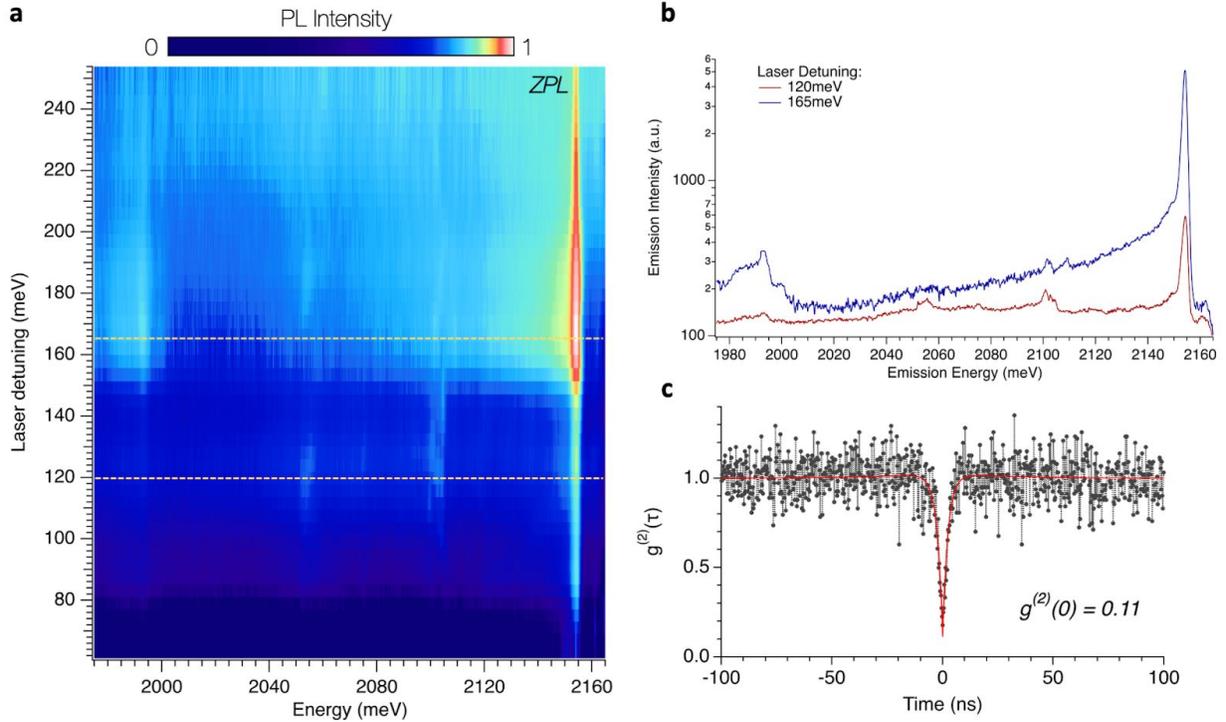

**Figure 2: Absorption spectrum of quantum emitter *E2* in hBN: a -** PLE map obtained by measuring the emission spectra at different excitation wavelengths with steps of 1 nm. Spectra are normalized according to the laser power at different wavelengths and intensity is plotted in logarithmic color scale. Laser detuning ($\omega_i$-$\omega_{ZPL}$) is calculated by fitting the center of the ZPL for each spectra. A clear increase in the absorption occurs for detuning above 150 meV. **b -** Examples of emission spectra, plotted in logarithmic scale, for excitation above (blue line) and below (red line) the detuning value of 150 meV, as indicated by dashed lines in (a) **c -** Room-temperature second-order autocorrelation function (black dots) and fitting function (red line).

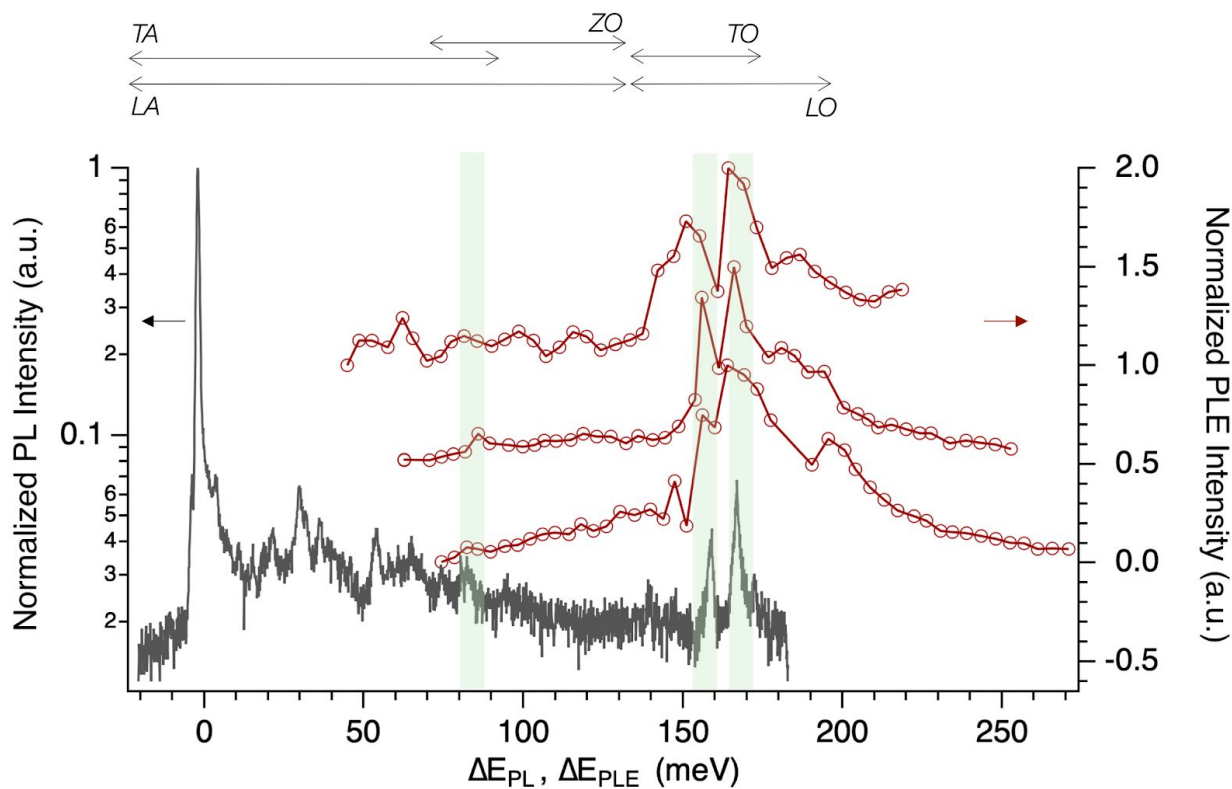

**Figure 3: Phonon resonances of emission and absorption spectra.** The intensity profiles of the ZPL extracted from PLE maps are plotted with red dots as a function of the laser detuning ($\Delta E_{PLE}$) for emitters *E1 - E3*. PLE spectra are plotted with an offset of 0.5 for the sake of clarity. Solid red lines are a guide for the eye. These specific emitters have ZPL energies between 2.1 and 2.16 eV. The emission spectrum for one of these emitters is illustrated with black line in logarithmic scale and mirrored with respect to the ZPL energy ($\Delta E_{PL}$). Other emission spectra are shown in Figure 1b and Figure 4S. Several resonances are visible and are associated to the different phonon modes as highlighted with light green areas. The approximate energy bands covered by LO and LA phonon modes are defined by horizontal arrows according to Ref.[27]. PLE spectra covering larger values of the detuning are reported in Figure 3S of the Supplementary Material.

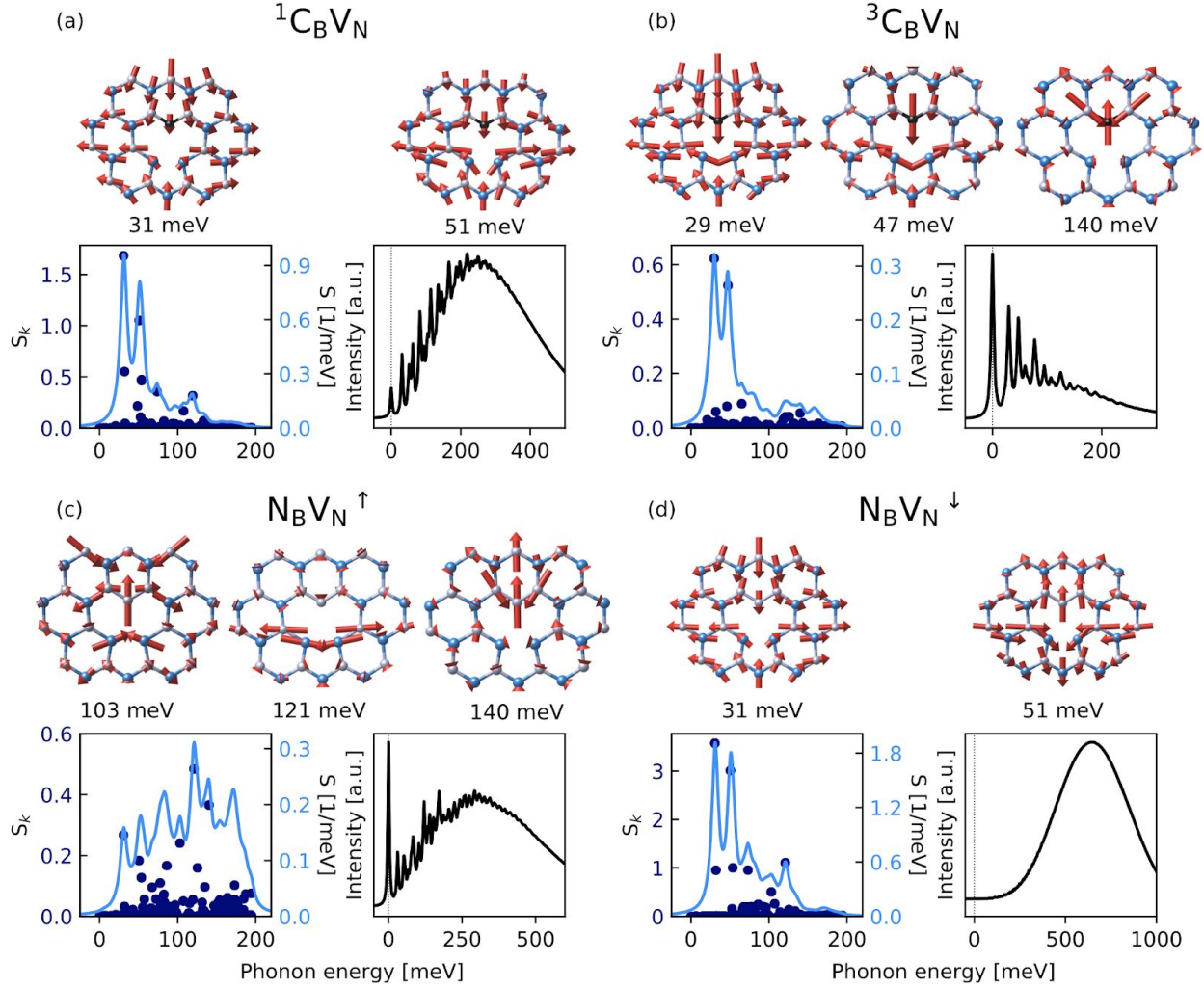

**Figure 4: First-principles calculations of defect-phonon coupling: a-d:** Calculated partial Huang-Rhys factor $S_k$ for phonon mode k, (using the methods outlined in Ref.[35]) along with the predicted spectra relative to the zero phonon line calculated for $C_BV_N$ and $N_BV_N$ defects modeled in multilayer hBN. We consider transitions in $C_BV_N$ involving a ground state spin singlet (a) and triplet (b), along with the spin-majority (c) and spin-minority (d) transitions involving the defect $N_BV_N$. The phonon mode displacement vectors for relevant strongly-coupling phonon modes are also visualized for regions near the defect. The transitions in (a), (b) and (d) each show similar mode-coupling profiles, notably strong coupling to two particular phonon modes around 30 and 50 meV. Additional phonon coupling is found in all three transitions for phonon modes between ~100 and 150 meV, which may be relevant for the large coupling seen here experimentally. These transitions however differ significantly in their efficiency, as described by the predicted luminescence lineshape, which is only predicted to be efficient into the zero phonon line for $^3C_BV_N$. The overall coupling profile is qualitatively different in the spin majority transition for $N_BV_N$ (panel c), where the coupling to phonons occurs across the phonon energy spectrum.

## METHODS

**Sample Fabrication**

The hBN thin-films used in these experiments was fabricated via low pressure chemical vapor deposition and transferred according to the method explained in Ref [10]. Briefly, hBN was grown on a copper catalyst, using ammonia borane as a precursor. The furnace temperature was set to 1030 °C and kept at a pressure of 2 torr under a 5% $H_2$/Ar atmosphere during the growth. The as-grown films were then transferred from copper to Si/$SiO_2$ (285 nm) substrate via a PMMA assisted wet transfer process. The polymer layer was then removed by soaking the sample in warm acetone (~50 °C) overnight, before further cleaning by exposure to UV-Ozone environment for 20 minutes, and annealing in air at 550 °C for 2 hours.

**Optical Characterization**

Low temperature measurements are run in a closed-cycle cryostat from Montana Instruments at T= 4 K. Tunable excitation is achieved by narrow-band (< 1 MHz) laser M2 SolsTiS EMM. PLE maps are obtained recording spectra at different wavelength with 1nm steps. Laser power for each excitation frequency is measured before the objective lens. Spectra are recorded with a IsoPlane SCT 320 spectrometer from Princeton Instrument. Emission counts are measured by APD modules from Excelitas (SPCM-AQRH). Antibunching measurements are fitted with second-order autocorrelation function $g^{(2)}(t) = 1 - Ae^{-\frac{t}{|\tau_1|}} + Be^{-\frac{t}{|\tau_2|}}$, where parameters $\tau_1$ and $\tau_2$ are the lifetimes of the excited and a possible metastable state, respectively, while $A$ and $B$ are fitting parameters.

**Computational details**

All calculations were performed using density functional theory within JDFTx.[33] The multilayer hBN system was described by placing the defect in a 7x7 supercell of hBN unit cells in the in-plane direction with an hBN layer both above and below the defect-hosting layer. Calculations were performed by sampling the Γ point using a 25 Hartree energy cutoff for a plane-wave basis set. The electron-nuclear interaction is described using ultrasoft pseudopotentials[38] and the exchange-correlation interaction is described by the PBE exchange correlation functional.[39] To capture the van der Waals nature of the out-of-plane interactions, the Grimme DFT + D2 scheme is used.[40] In our spin-polarized calculations, $N_BV_N$ is found to be a spin triplet, for which we evaluate transitions involving both the spin majority and spin minority defect states. We also evaluate spin majority transitions in the spin triplet and singlet ground states of $C_BV_N$, the latter of which is found to be the energetically preferred state. In all cases, we excite the system and ionically relax the system using ΔSCF before calculating the coupling with the phonon modes of the relaxed ground state geometry. The relaxation is confined to in-plane displacements. To characterize the optical properties of the atomic defects, we calculate the ab-initio emission line shapes using the method outlined in Ref.[35].

**ACKNOWLEDGMENT**

This work (H.M., C.C., P.N., D.E.) was supported by the Army Research Office (ARO) Multidisciplinary University Research Initiative (MURI) program, grant no. W911NF-18-1-0431. G.G. acknowledges support from the Graduate Center of the City University of New York (CUNY) through start-up funding. H.M. acknowledges support by Samsung Scholarship. J. F. acknowledges financial support from the Deutsche Forschungsgemeinschaft under Contract No. FL 997/1-1. P.N. is a Moore Inventor Fellow supported in part by the Gordon and Betty Moore Foundation.